\documentclass[journal=jpcafh,manuscript=article]{achemso}
\usepackage{amsmath}
\usepackage{amsmath,amssymb,amsfonts,latexsym,graphicx}
\usepackage{geometry}
\usepackage{indentfirst}
\usepackage{fancyhdr}
\usepackage{subfigure}
\usepackage{algorithm}
\usepackage{algorithmic}
\usepackage{titletoc}
\usepackage{titlesec}
\usepackage{booktabs}
\usepackage{threeparttable}
\usepackage{multirow}
\usepackage{color}
\usepackage{rotating}
\usepackage[version=3]{mhchem} 
\setkeys{acs}{maxauthors = 0} 

\title{Accelerating Fock build via hybrid analytical-numerical integration}

\author{Yong Zhang}
\affiliation{Qingdao Institute for Theoretical and Computational Sciences, College of Chemistry and Chemical Engineering,
Shandong University, Qingdao 266237, P. R. China}

\author{Rongding Lei}
\author{Bingbing Suo}\email{bsuo@nwu.edu.cn}
\affiliation{Shaanxi Key Laboratory for Theoretical Physics Frontiers, Institute of Modern Physics, Northwest University, Xi'an 710127, P. R. China}

\author{Wenjian Liu}
\email{liuwj@sdu.edu.cn}
\affiliation{Qingdao Institute for Theoretical and Computational Sciences, College of Chemistry and Chemical Engineering,
Shandong University, Qingdao 266237, P. R. China}

\keywords{Hatree-Fock; Density functional theory; Fock-Like Matrix; TD-DFT}

\begin{document}

\begin{abstract}
A hybrid analytical-numerical integration scheme is introduced to accelerate the Fock build in
self-consistent field (SCF) and time-dependent density functional theory (TDDFT) calculations.
To evaluate the Coulomb matrix $\mathbf{J}[\mathbf{D}]$, the (induced) density matrix $\mathbf{D}$
is first decomposed into two parts, the superposition of atomic density matrices $\mathbf{D}^A_{\oplus}$
and the rest $\mathbf{D}^R=\mathbf{D}-\mathbf{D}^A_{\oplus}$. While $\mathbf{J}[\mathbf{D}^A_{\oplus}]$ 
is evaluated analytically, $\mathbf{J}[\mathbf{D}^R]$ is evaluated fully numerically [with the
multipole expansion of the Coulomb potential (MECP)]
during the SCF iterations. 
Upon convergence, $\mathbf{D}^R$ is further split 
into those of near ($\mathbf{D}^{RC}$) and distant ($\mathbf{D}^{RL}$) atomic orbital (AO) pairs,
such that $\mathbf{J}[\mathbf{D}^{RC}]$ and $\mathbf{J}[\mathbf{D}^{RL}]$ are evaluated semi-numerically and fully numerically (with MECP). 
Such a hybrid $\mathbf{J}$-build is dubbed `analytic-MECP' (aMECP).
Likewise, the analytic evaluation of $\mathbf{K}[\mathbf{D}^A_{\oplus}]$ and semi-numerical evaluation of $\mathbf{K}[\mathbf{D}^R]$
are also invoked for the construction of the exchange matrix $\mathbf{K}[\mathbf{D}]$ during the SCF iterations. 
The chain-of-spheres (COSX) algorithm  [Chem. Phys. 356, 98 (2009]) is employed for $\mathbf{K}[\mathbf{D}^R]$ 
but with a revised construction of the S-junctions for overlap AO pairs. To distinguish from the original COSX algorithm 
(which does not involve the partition of the density matrix $\mathbf{D}$), we denote the presently revised variant as COSx.
Upon convergence, $\mathbf{D}^R$ is further split into those of near ($\mathbf{D}^{RC}$) and distant ($\mathbf{D}^{RL}$) AO pairs
followed by a re-scaling, leading to $\tilde{\mathbf{D}}^{RC}$ and $\tilde{\mathbf{D}}^{RL}$, respectively. 
$\mathbf{K}[\tilde{\mathbf{D}}^{RC}]$ and $\mathbf{K}[\tilde{\mathbf{D}}^{RL}]$ 
are then evaluated analytically and semi-numerically (with COSx), respectively. 
Such a hybrid $\mathbf{K}$-build is dubbed `analytic-COSx' (aCOSx). Extensive numerical experimentations reveal
that the combination of aMECP and aCOSx is highly accurate for ground state SCF calculations ($<1 \mu\mathrm{E}_h/\mathrm{atom}$ error in energy)
and is particularly efficient for calculations of large molecules with extended basis sets. 
As for TDDFT excitation energies, a medium grid for MECP and a coarse grid for COSx are already sufficient.
\end{abstract}

\section{Introduction}
An efficient Fock build is the prerequisite for electronic structure calculations of large molecular systems\cite{OchsenfeldRev2013}.
In the atomic orbital (AO) representation, the Fock matrix $\mathbf{F}$ is composed of a one-electron term $\mathbf{h}$
and two two-electron terms,
the Coulomb ($\mathbf{J}$) and exchange ($\mathbf{K}$) matrices. The latter two arise from contractions between the
two-electron repulsion integrals (ERI) and the elements of a given density matrix $\mathbf{D}$, viz.,
\begin{align}
J_{\mu\nu}& = \sum_{\lambda\sigma}(\mu \nu|\lambda \sigma) D_{\sigma\lambda},  \label{eqn:jmat} \\
K_{\mu\nu} &= \sum_{\lambda\sigma} (\mu \sigma|\lambda\nu )D_{\sigma\lambda},  \label{eqn:kmat}
\end{align}
where the ERIs have been written in the Mulliken notation and the density matrix can be that
in Hartree-Fock (HF), Kohn-Sham (KS), or time-dependent density functional (TDDFT)\cite{Casida2009TDDFT} theories.
While the ERIs can be evaluated analytically when Gaussian functions are taken as the AOs,
the formally quartic scaling places a bottleneck on both computational cost and memory footprint.
The very first key for reducing the computational cost and memory requirement
is the use of integral upper bounds\cite{IntBounds1973,IntBounds1994,IntBounds2012,IntBounds2019}
to pre-screen the ERIs on one hand and recalculate them whenever needed on the other hand.
Such integral-direct self-consistent field (SCF) has an asymptotically quadratic scaling,
due to the fact that the number of numerically significant AO pairs scales linearly with molecular size
(characterized by the number of atoms)\cite{NoN21973,SchwarzBond1989}.
Since the pre-screening tends to pick up different batches of the ERIs for the Coulomb and exchange interactions,
it is advantageous to construct them separately to break the quadratic wall\cite{SeparateJK1991}.
For the former, the resolution of the identity (RI; also called density fitting)\cite{RI-1,WhittenRI1973,RI-2,DunlapRI1979,VahtrasRI1993,WeigendRIJBasis1997,WeigendRICBasis2002,NeeseJCC2003RI1} turns out to be very effective for medium-sized systems (up to a few hundred atoms).
Therein, a set of fitting functions is invoked to approximate the AO products, such that the expensive four-center ERIs
can be assembled in terms of two- and three-center ERIs. Although the scaling with molecular size is cubic here  (due
to the matrix inversion needed to calculate the expansion coefficients), the prefactor is very small.
Given well-trained fitting functions\cite{eichkorn1995auxiliary,WeigendRIJBasis1997,DEF2TVZP,Weigend2002RI},
very high accuracy can be achieved\cite{EichkornRIJ1995, WeigendRIJBasis1997, SkylarisRI2000}.
Yet, some care should be taken to avoid unphysical situations (e.g., zero or even negative fitted densities)
when using the fitting functions optimized for ground state calculations
in otherwise excited state calculations\cite{van2003application,BDFTDDFT2004}.
Such problems are less prone in the Cholesky decomposition (CD) approach\cite{CD-1,CD-2,CD-3,CD-4,CD-Rev},
where the fitting functions (Cholesky vectors) are generated on-the-fly with well-controlled accuracy, albeit at the price of some
computational overhead. For very large systems, the charge distributions (AO pairs) can be decomposed
into overlapping and well-separated ones, thereby splitting the Coulomb potential $V_c(\vec{r})$ into
near- and far-field contributions. Although the former must be calculated directly with the RI/CD-ERIs,
the latter can be evaluated by means of multipole expansions\cite{white1994derivation,CFMM1994,GCFMM1995,GvFMM1996,QCTC1996,sierka2003fast,BFMM2004,Lazarski2015density},
thereby achieving a linear-scaling construction of the Coulomb matrix. Since the near-field contributions
scale linearly with molecular size (due to the local nature of the basis functions),
the linear-scaling construction of the Coulomb matrix can be achieved even for systems with a
delocalized electronic structure (i.e., with a non-sparse density matrix).

The situation is different for the exchange matrix: a linear-scaling construction can only be
achieved for gapped systems featuring localized electronic structures, because only in such cases
the density matrix elements $D_{\sigma\lambda}$ tend to decay exponentially with respect to spatial separations of the basis
functions $\sigma$ and $\lambda$\cite{kohn1995density,Schwegler1997ONX}. It follows that the four functions in the exchange ERI $(\mu\sigma|\lambda\nu)$
must then be close to each other in space to have a significant contribution. Consequently, the number of numerically significant
ERIs for the exchange interaction scales linearly with molecular size. However, the pre-screening step still scales quadratically.
Several algorithms\cite{schwegler1996linear,MultipoleX1996,Schwegler1997ONX,Link1998,Schwegler2000SONX,Ochsenfeld2013ExtLinK} have been designed to resolve this.
Yet, their performances deteriorate for extended basis sets (especially in the presence of diffuse functions), due to reduced
sparsity of the density matrix. The problem can largely be resolved by introducing an auxiliary density matrix and using a
density functional correction for the difference between the true and auxiliary density matrices\cite{Guidon2010ADMM}.
The exchange matrix can also be constructed with the RI/CD technqiues\cite{Fruchtl1997RIK,Weigend2002RI,weigend2008exchange,CD-Rev}
but which become effective only with local fittings\cite{Polly2004LocalFit,Sodt2008ARIK,AquilanteaCD2007,reine2008variational,AquilanteaCD2009,Merlot2013PARIK}.
Moreover, it is possible to further accelerate the construction of the exchange matrix by
evaluating a large portion of the exchange ERIs with multipole expansions\cite{MultipoleX1998,MultipoleX1999}.

Apart from the above analytical evaluations, the Coulomb and exchange matrices can also be constructed semi-numerically\cite{Friesner1985,Neese2009COSX1,Neese2011COSX2,Neese2021COSX3,NakaimCOSX2017,semiX2020,semiX2021}
and even fully numerically\cite{NumVc1994,NumVc1995,Delley1990MECP,Liu1997-BDF,Liu2003-BDF}.
Among these, the semi-numerical change-of-spheres algorithm for the exchange build (COSX) \cite{Neese2009COSX1,Neese2011COSX2,Neese2021COSX3,NakaimCOSX2017}
is particularly advantageous for extended basis sets with functions of high angular momenta and
meanwhile has a good balance between performance and accuracy\cite{Rebolini2016RIKCOSX,Laqua2020GPUCOSX}, whereas
the fully numerical multipole expansion of the Coulomb potential (MECP)\cite{Delley1990MECP,Liu1997-BDF,Liu2003-BDF}
is most efficient for the Coulomb build, for it is independent of basis set size and electron kinematics (nonrelativistic or relativistic).
Yet, the errors resulting from the use of finite grids are often case dependent, which not only affect the calculated energies and gradients,
but may also affect iteration processes. This is precisely the issue we want to handle in this work.
It will be shown that the evaluation of the $\mathbf{J}[\mathbf{D}]$ matrix can be facilitated
by decomposing the density matrix $\mathbf{D}$ into three portions, i.e., the superposition of atomic density matrices ($\mathbf{D}^A_{\oplus}$),
the portions for near ($\mathbf{D}^{RC}$) and distant ($\mathbf{D}^{RL}=\mathbf{D}^R-\mathbf{D}^{RC}$;
$\mathbf{D}^R=\mathbf{D}-\mathbf{D}^A_{\oplus}$) AO pairs, such that
$\mathbf{J}[\mathbf{D}^A_{\oplus}]$ can readily be evaluated analytically, whereas $\mathbf{J}[\mathbf{D}^{RC}]$ and $\mathbf{J}[\mathbf{D}^{RL}]$
can be evaluated to sufficient accuracy with semi-numerical integration and MECP, respectively. As a matter of fact,
the latter two can be combined (i.e., $\mathbf{J}[\mathbf{D}^R]$) and evaluated with MECP during the iterations. That is,
the separate treatment of $\mathbf{J}[\mathbf{D}^{RC}]$ and $\mathbf{J}[\mathbf{D}^{RL}]$ is performed only once upon convergence, so as
to avoid repeated evaluations of the relatively expensive $\mathbf{J}[\mathbf{D}^{RC}]$. The so-defined hybrid $\mathbf{J}$-build is dubbed analytical-MECP (aMECP).
Likewise, the analytic evaluation of $K[\mathbf{D}^A_{\oplus}]$ and semi-numerical evaluation of $K[\mathbf{D}^R]$
are also invoked for the construction of the exchange matrix $K[\mathbf{D}]$ during the SCF iterations. 
The COSX algorithm\cite{Neese2009COSX1} is employed for $K[\mathbf{D}^R]$ 
but with a revised construction of the S-junctions for overlapping AO pairs. To distinguish from the original COSX algorithm 
(which does not involve the partition of the density matrix $\mathbf{D}$), we denote the presently revised variant as COSx.
Upon convergence, $\mathbf{D}^R$ is further split into re-scaled near ($\tilde{\mathbf{D}}^{RC}$) and distant ($\tilde{\mathbf{D}}^{RL}$) AO pairs.
The former is evaluated analytically, whereas the latter is evaluated via COSx. 
Such a hybrid $\mathbf{K}$-build is dubbed `analytic-COSx' (aCOSx).
The detail of aMECP and aCOSx  will be presented in Sec. \ref{algorithms} and benchmarked in Sec. \ref{Results}.
Some concluding remarks will be provided in Sec. \ref{Conclusion} to close the presentation.

\section{Methodologies}\label{algorithms}
\subsection{aMECP}
For a semi-numerical evaluation of the Coulomb matrix elements, we rewrite Eq. \eqref{eqn:jmat} as
\begin{align}
J_{\mu\nu}[\mathbf{D}]&=\int \chi_{\mu}(\vec{r})\chi_{\nu}(\vec{r})V_{c}[\rho](\vec{r}) d^3r\\
&\approx \sum_{g}^{N_g} w(\vec{r}_g) \chi_{\mu}(\vec{r}_g)V_{c}[\rho](\vec{r}_g)\chi_{\nu}(\vec{r}_g),\label{eqn:Jmat}\\
V_{c}[\rho](\vec{r}_g)&=\int \frac{\rho(\vec{r})}{|\vec{r}_g-\vec{r}|}d^3r,\quad
\rho(\vec{r}) = \sum_{\lambda\sigma} \chi_\lambda(\vec{r})\chi_{\sigma}(\vec{r})D_{\sigma\lambda} \\
&=\sum_{\lambda\sigma}  A^g_{\lambda\sigma} D_{\sigma\lambda},\label{eqn:Vamat}\\
A_{\lambda\sigma}^g&=\int \frac{\chi_\lambda(\vec{r})\chi_\sigma(\vec{r})}{|\vec{r}_g-\vec{r}|}d^3r, \label{eqn:amat}
\end{align}
where $\{\vec{r}_g\}_{g=1}^{N_g}$ and $\{w(\vec{r}_g)\}_{g=1}^{N_g}$ represent the integration grid points and corresponding weights, respectively.
The $\mathbf{A}^g$ matrix \eqref{eqn:amat} for each grid point involves 3-center-1-electron (3c1e) Coulomb integrals and
can be evaluated analytically in the same way as that for the attraction potential of a nucleus (see Sec. S1 in the Supporting Information for a special treatment
of such 3c1e integrals over basis functions of summed angular momenta up to 4).
However, since the number of grid points is typically thousands of times the number of atoms,
the computational of all $\mathbf{A}^g$ has a quadratic scaling with a very large prefactor.
Here, we adopt a hybrid analytical-numerical scheme by first decomposing the density matrix as
\begin{align}
\mathbf{D}=\mathbf{D}^A_{\oplus}+\mathbf{D}^R, \quad \mathbf{D}^R=\mathbf{D}^{RC}+\mathbf{D}^{RL},
\end{align}
and hence the Coulomb matrix as
\begin{align}
\mathbf{J}[\mathbf{D}]=\mathbf{J}[\mathbf{D}^A_{\oplus}]+\mathbf{J}[\mathbf{D}^R],
\quad \mathbf{J}[\mathbf{D}^R]=\mathbf{J}[\mathbf{D}^{RC}] +\mathbf{J}[\mathbf{D}^{RL}].
\end{align}
Here, $\mathbf{D}^A_{\oplus}$ is the superposition of atomic density matrices obtained from calculations of
spherical and unpolarized atomic configurations, and $\mathbf{D}^{RC}$ represents
the portion of $\mathbf{D}$ for near AO pairs, viz.,
\begin{align}
D^{RC}_{\sigma\lambda} = \left \{
\begin{array}{cr}
D^R_{\sigma\lambda} & [R_{\sigma\lambda}<R^{cut}_{\sigma}+R^{cut}_{\lambda}] \cap [D^{R}_{\sigma\lambda}>\eta_{D}] \\
0 & [R_{\sigma\lambda} \geq R^{cut}_ {\sigma}+R^{cut} _{\lambda}] \cap [D^{R}_{\sigma\lambda} \leq \eta_{D}],
\end{array}
\right.
\label{J-AOpair}
\end{align}
where $\eta_D$ ($=10^{-8}$) is a predefined threshold,
$R_{\sigma\lambda}$ is the distance between atomic centers of two primitive Gaussian functions
$\chi_{\sigma}$ and $\chi_{\lambda}$, whereas the cutoff radius $R^{cut}_{\sigma}$ of function $\chi_{\sigma}$ is
determined by the following radial integration
\begin{align}
\int_{r=R^{cut}_{\sigma}}^\infty R^{2}_{\sigma}(r) r^2 dr=\eta_b.  \label{eqn:bfradius1}
\end{align}
in which $R_\sigma(r)$ is the radical part of a primitive GTO. The parameter $\eta_b$ is usually to be set to $0.30$ in energy calculations but to $0.35$ in gradient calculations.
Although the semi-numerical evaluation of $\mathbf{J}[\mathbf{D}^{RC}]$ (cf.
Eqs. \eqref{eqn:Jmat} and \eqref{eqn:Vamat}) has a quadratic scaling, the band structure of $\mathbf{D}^{RC}$
ensures a small prefactor. This holds also for the analytical evaluation of $\mathbf{J}[\mathbf{D}^A_{\oplus}]$ due to the
diagonal-in-atom structure of $\mathbf{D}^A_{\oplus}$.
In particular, thanks to the spherical averaging, each atomic density matrix has vanishing elements between basis functions of different angular momenta
and between functions that have no contributions to the occupied atomic orbitals. As for the potential due to the residual density matrix $\mathbf{D}^{RL}$,
which is small in magnitude but may spread over the whole system,
we adopt the MECP approach\cite{Delley1990MECP,Liu1997-BDF,Liu2003-BDF}, where
the residual density $\rho^{RL}(\vec{r})$ is first partitioned into a sum of one-centered contributions via a normalized partition function\cite{BeckePartition},
\begin{align}
\rho^{RL}(\vec{r})=\sum_{\lambda\sigma}\chi_{\lambda}(\vec{r})\chi_{\sigma}(\vec{r})D^{RL}_{\sigma\lambda}=\sum_A P_A(\vec{r})\rho^{RL}(\vec{r})=\sum_A\rho^{RL}_A(\vec{r}).\label{AtomDen}
\end{align}
The one-centered densities are then projected onto their multipole components,
\begin{align}
\rho^{RL}_{Alm}(r_A) &=\int_{|\vec{r}-\vec{R}_A|=r_A} Y_{lm}(\hat{\vec{r}}_A)\rho^{RL}_A(\vec{r})d\Omega,\quad r_A=|\vec{r}-\vec{R}_A|,
\end{align}
so as to approximate the residual density as
\begin{align}
\rho^{RL}(\vec{r})=\sum_A\sum_{l=0}^{L_{\max}}\sum_{m=-l}^lY_{lm}(\hat{\vec{r}}_A)\rho^{RL}_{Alm}(r_A).\label{TotalDen}
\end{align}
Further using the multipole expansion of $1/|\vec{r}_1-\vec{r}_2|$, the Coulomb potential can be calculated as
\begin{align}
V_c [\rho^{RL}] (\vec{r}) &= \sum_A \sum_{l=0}^{L_{max}}\sum_{m=-l}^{l} \frac{4 \pi} { 2l+1 } Y_{lm}(\hat{\vec{r}}_A) V^{RL}_{Alm}(r_A),\label{VcMECP}\\
V^{RL}_{Alm}( r_A) &= \frac{1} { r_A^{l+1} } \int_0^{r_A}\rho^{RL}_{Alm} (s) s^{l+2} ds + r_A^l \int_{r_A}^{\infty} \rho^{RL}_{Alm} (s)s^{1-l} ds.\label{Valm}
\end{align}
To facilitate the calculations of the right hand sides of Eqs. \eqref{Valm} and \eqref{VcMECP},
the tabulated $r_A^2 \rho^{RL}_{Alm}(r_A)$ and $V^{RL}_{Alm}(r_A)$ are fitted with cubic splines.
Plugging $V_c [\rho^{RL}] (\vec{r}_g)$ into Eq. \eqref{eqn:Jmat} gives rise to $J_{\mu\nu}[\mathbf{D}^{RL}]$.
Eqs. \eqref{AtomDen}-\eqref{TotalDen} and \eqref{Valm} all scale linearly, but Eq. \eqref{VcMECP}
scales quadratically  with molecular size. However, the prefactor of the latter is very small, due to the short-range nature of the numerical functions
$\{V^{RL}_{Alm}(\cdot)\}$ for angular momenta higher than 2.  Compared to other density fitting approaches
\cite{RI-1,WhittenRI1973,RI-2,DunlapRI1979,VahtrasRI1993,WeigendRIJBasis1997,WeigendRICBasis2002,NeeseJCC2003RI1},
the MECP scheme 
amounts to using saturated fitting functions of angular momenta up to $L_{max}$, but without
explicit fitting functions. It also deserves to be mentioned that the present extraction of $\mathbf{D}^A_{\oplus}$
from the full density matrix $\mathbf{D}$ is more natural and more accurate than the introduction\cite{NumVc1995} of a crude model density matrix
for the purpose of curing the numerical errors arising from the long-range part
of the numerically evaluated Coulomb potential\cite{NumVc1994}.

As a matter of fact, the relatively expensive semi-numerical evaluation of $\mathbf{J}[\mathbf{D}^{RC}]$ \eqref{J-AOpair} need not be performed in each SCF iteration. Instead, the use of MECP for $\mathbf{J}[\mathbf{D}-\mathbf{D}^A_{\oplus}]$ is sufficient during the iterations. As such, the separate treatment of $\mathbf{J}[\mathbf{D}^{RC}]$ and $\mathbf{J}[\mathbf{D}^{RL}]$ needs to be performed only once in the end of iterations, which is necessary to ensure the accuracy.
It also deserves to be mentioned that MECP itself is sufficiently accurate for the induced Coulomb potential entering TDDFT calculations. 

\subsection{aCOSx}\label{aCOSXSec}
For a semi-numerical evaluation of the exchange matrix, Eq. \eqref{eqn:kmat} can be rewritten as\cite{Neese2009COSX1} 
\begin{align}
K_{\mu\nu}&\approx \frac{1}{2}(\tilde{K}_{\mu\nu}+\tilde{K}_{\nu\mu}),\\
\tilde{K}_{\mu\nu}&=\sum_{g}^{N_g} w(\vec{r}_g) \chi_{\mu}(\vec{r}_g)\chi_{\sigma}(\vec{r}_g)A^g_{\lambda\nu}D_{\sigma\lambda}=
 (\mathbf{X}\mathbf{G}^\dag)_{\mu\nu},\\
X_{\mu g} &= w_g^{1/2}\chi_{\mu}(\vec{r}_g),\quad F_{\lambda g}=(\mathbf{D}\mathbf{X})_{\lambda g},\quad G_{\nu g}=\sum_{\lambda}A^g_{\nu\lambda}F_{\lambda g}.
\end{align}
It can be shown\cite{Neese2009COSX1} that 
even the steepest scaling intermediate $G_{\nu g}$ has a very favourable 
scaling with the maximum angular momentum in the basis set, as compared with the analytical evaluation of $K_{\mu\nu}$.
The numerical accuracy can further be improved by means of the overlap fitting\cite{Neese2011COSX2}, 
\begin{align}
\mathbf{K} &\approx \frac{1}{2}(\mathbf{S}\mathbf{S}_{\mathrm{num}}^{-1}\tilde{\mathbf{K}}+c.c.),
\end{align}
where $\mathbf{S}$ and $\mathbf{S}_{\mathrm{num}}$ ($=\mathbf{X}\mathbf{X}^\dag$) are the overlap matrices calculated analytically and numerically, respectively. 

For an efficient implementation of the above scheme, the COSX algorithm\cite{Neese2009COSX1} first determines
the radial extension of each shell of contracted basis functions and then establishes a list
of partners (S-junction) for each basis function according to an overlap criterion. 
A second list of parters (P-junction) is further constructed for each basis function according to a cutoff criterion for 
the density matrix elements. The length of the S-junction for every basis function approaches constant due to
the short-range nature of the basis function, whereas the length of the 
P-junction is determined by the degree of localization of the electronic structure (i.e., sparsity of the density matrix).
Pictorially, two basis functions $\chi_{\mu}$ and $\chi_{\nu}$ are confined in their spheres $\mathcal{S}_{\mu}$ and $\mathcal{S}_{\nu}$,
respectively, which are  
connected to a number of spheres $\{\mathcal{S}_{\sigma}\}$ of $\{\chi_{\sigma}\}$ and
$\{\mathcal{S}_{\lambda}\}$  of $\{\chi_{\lambda}\}$, respectively. The latter spheres
are further connected, thereby forming chains of (atom-centered and connected) spheres for the basis pair $\chi_{\mu}\chi_{\nu}$
in the exchange ERIs $(\mu\sigma|\lambda\nu)$. In the aCOSx approach presented here, instead of determining the radial extensions of shells of \emph{contracted} basis functions according to their absolute values in a point-wise manner, 
the radial extensions $\{R_{\mu}\}$ of \emph{primitive} basis functions $\{\chi_{\mu}\}$ are estimated simply according to the integral 
defined in Eq. \eqref{eqn:bfradius1}
with $\eta_b=0.001$. Two primitive functions $\chi_{\mu}$ ($\chi_{\nu}$) and $\chi_{\sigma}$ 
($\chi_{\lambda}$) are connected if the sum of their radii is smaller than
the distance between the atoms to which they belong.
The functions $\chi_{\sigma}$ and $\chi_{\lambda}$ in the exchange ERIs $(\mu\sigma|\lambda\nu)$
are regarded as connected only if the density matrix element $D_{\sigma \lambda}$ 
is larger than a preset threshold $\delta_D$ ($= 10^{-8}$). Numerical experimentations reveal that the  aCOSx
is more effective than the original COSX in reducing the number of numerically significant 3c1e integrals $A_{\lambda\nu}^g$,
albeit with a somewhat larger memory requirement.    

The number of grid points is a crucial factor for the efficiency of aCOSx. 
After extensive experimentations, we propose the following rule for the number of Gauss-Chebeshev type of radial grid points, 
\begin{align}
N_R(r,Z,K,L)=5 r+N_q(Z)+5 K+15L,\quad K, L \in [0, 9],  \label{RadialGrid}
\end{align}
where $r$ is the row of the period table to which the atom belongs, whereas $N_q$ is function of nuclear charge $Z$ (see Table \ref{tbl:COSXGRID}).
Since it is natural\cite{Gill1993SG1} to use different Lebedev angular grids\cite{Lebedev1975,Lebedev1976} for different regions in space, 
the atomic radius $R$ is decomposed into seven intervals $[a_{i-1}R, a_i R)$, 
with $a_0=0$ and $a_i$ ($i\in[1,6]$) given in Table \ref{tbl:radregion}. The numbers of angular grid points 
for different radial intervals are controlled by the parameter $M+\Delta M$: 
$M$ ($\ge 1$) in Table \ref{tbl:anggridbyM} tells the number of angular grid points for the fourth interval $[a_3 R, a_4 R)$,
whereas $\Delta M$ in Table \ref{tbl:anggridadjust} tells the reduced number of angular grid points for other intervals.
Overall, the size of the grid in the aCOSx calculations can be characterized by function $G_{KLM}$. For instances, $G_{002}$ is typically used in SCF and TDDFT calculations. The so-generated grids are finally pruned\cite{adapgrid1998,adapgrid2004}.

\begin{table}
\centering
\caption{Function $N_q(Z)$ in Eq. \eqref{RadialGrid} }
\label{tbl:COSXGRID}
\begin{threeparttable}
\begin{tabular}{ccccccc}
\hline\hline
 atom  & \ce{H}-\ce{N}e & \ce{N}a-\ce{Kr} & \ce{Rb}-\ce{Xe} & \ce{Cs}-\ce{Yb} &
  \ce{Lu}-\ce{Hg} & \ce{Tl}- \\
  \hline
 $N_q(Z)$ & 10 & 15 & 20 &25 &30 &40 \\
 \hline
\end{tabular}
\end{threeparttable}
\end{table}

\begin{table}
\centering
\caption{Parameters for defining radial intervals $[a_{i-1}R, a_i R)$ ($a_0=0$)}
\label{tbl:radregion}
\begin{threeparttable}
\begin{tabular}{lcccccc}
\hline
\hline
                 &          \multicolumn{6}{c}{ $a_i$} \\
               \cline{2-7}
atom                     & $a_1$ & $a_2$ & $a_3$ & $a_4$ & $a_5$ & $a_6$ \\
\hline
\ce{H}-\ce{He}  & 0.25   & 0.50   & 1.00 & 4.50 & 7.80 & 10.00   \\
\ce{Li}-\ce{Ne} & 0.17   & 0.50   & 0.90 & 3.50 & 7.80 & 9.00     \\
\ce{Na}-\ce{Ar} & 0.10   & 0.40   & 0.80 & 2.50 & 5.00 & 7.50     \\
\ce{K}-                  & 0.02   & 0.10   & 0.20 & 3.50 & 7.00 & 9.00     \\
\hline
\end{tabular}
\end{threeparttable}
\end{table}

\begin{table}
\caption{Correspondence between $M$ in $G_{KLM}$ and number of Lebedev angular grids}
\begin{threeparttable}
\begin{tabular}{lccccccccccccccccc}
\hline\hline
                &          \multicolumn{14}{c}{ $M$} \\
                 \cline{2-15}
  atom           & -4 & -3 & -2 & -1 & 0 & 1 &  2 & 3  & 4 & 5 & 6 & 7 & 8 & 9     \\
\hline
\ce{H} -\ce{Pd} & 6 & 6 &  6 & 14 & 26  & 38  & 50  & 110 & 194 & 302 & 434 & 590 &  770 & 974  \\
\ce{Ag}-\ce{Yb} & 6 & 6 & 14 & 26 & 38  & 50  & 110 & 194 & 302 & 434 & 590 & 770 &  974 & 1202 \\
\ce{Lu}-        & 6 & 14& 26 & 38 & 50  & 110 & 194 & 302 & 434 & 590 & 770 & 974 & 1202 & 1454 \\
 \hline
\end{tabular}
\end{threeparttable}\label{tbl:anggridbyM}
\end{table}
\begin{table}
\caption{Parameter $\Delta M$ for adjusting the number of Lebedev angular grids }
\begin{threeparttable}
\begin{tabular}{cccccccc}
\hline\hline
  interval    & [0, $a_1 R$) &  [$a_1 R$, $a_2 R$) & [$a_2 R$, $a_3 R$)  &  [$a_3 R$, $a_4 R$) &  [$a_4 R$, $a_5 R$) &  [$a_5 R$, $a_6 R$)  &   [$a_6 R$, $\infty$)     \\
\hline
   $\Delta M$  & -3 & -2 & -1 & 0 & -1 & -3 & -5 \\
 \hline
\end{tabular}
\end{threeparttable}\label{tbl:anggridadjust}
\end{table}

Another feature of the present aCOSx lies in that, like in aMECP, the density matrix $\mathbf{D}$ is decomposed as 
\begin{align}
\mathbf{D} = \mathbf{D}^A_{\oplus} + \mathbf{D}^R,
\end{align}
such that the exchange matrix is decomposed as 
\begin{align}
\mathbf{K}[\mathbf{D}] = \mathbf{K}[\mathbf{D}^A_{\oplus}] + \mathbf{K} [\mathbf{D}^{R}],\label{KDatom}
\end{align}
where $\mathbf{K}[\mathbf{D}^A_{\oplus}]$ needs to be calculated analytically only once, whereas
$\mathbf{K} [\mathbf{D}^{R}]$ is updated semi-numerically 
during the iterations with a smaller grid $G_{KLM}$. After convergence of the SCF iterations, the $D^R$ is first decomposed as
\begin{align}
\textbf{D}^R=\textbf{D}^{RC}+\textbf{D}^{RL},
\end{align}
where
\begin{align}
D^{RC}_{\mu\nu} = \left\{
\begin{array}{c}
D^R_{\mu\nu} \quad\quad [R_{\mu\nu} \le R^{cut}_{\mu} + R^{cut}_{\nu}] \cap [L(\chi_{p}) \le L^{A}_{OMax}(\chi_{p}), \quad p \in \mu,\nu],\\
0.
\end{array}
\right.
\end{align}
Here, $R^{cut}_\mu$ is the cutoff radius of a primitive basis function $\chi_\mu$ as defined in Eqn. \ref{eqn:bfradius1}. $L(\chi_p)$ represents the orbital angular moment of a basis function $\chi_p$, and $L^A_{OMax}(\chi_p)$ signifies the maximum angular moment of the occupied orbital of the atom A to which $\chi_p$ is associated. Then,  $\textbf{D}^{RC}$ and $\textbf{D}^{RL}$ are re-scaled as
\begin{align}
\tilde{\mathbf{D}}^{RC}=\mathbf{D}^{RC}+\frac{n^{RL}}{n^A}\mathbf{D}^A_{\oplus}, \label{eqn:DRC} \\
\tilde{\mathbf{D}}^{RL}=\mathbf{D}^{RL}-\frac{n^{RL}}{n^A}\mathbf{D}^A_{\oplus}, \label{eqn:DRL}
\end{align}
where
\begin{align}
n^{RL} = tr[\textbf{D}^{RL}\textbf{S}],\quad n^A=tr[\mathbf{D}^A_{\oplus}\mathbf{S}],
\end{align}
so as to render $tr[\tilde{\mathbf{D}}^{RL}\mathbf{S}]=0$, meaning that the distant AO pairs do not carry on net electrons. 
Finally, $\mathbf{K}[D^R]$ is calculated according to 
\begin{align}
\textbf{K}[{\mathbf{D}^R}] =\mathbf{K}[ \tilde{\textbf{D}}^{RC}] + \mathbf{K}[\tilde{\mathbf{D}}^{RL}],\label{KDAOpair}
\end{align}
where the first term is calculated analytically, whereas the second term is calculated by COSx with  a larger grid with $K+2$ and $M+1$. We will see in 
the benchmark calculations
that the aCOSx approach defined by Eqs. \eqref{KDatom} and \eqref{KDAOpair}
is more stable in accuracy than the original algorithm.\cite{Neese2009COSX1,Neese2011COSX2,Neese2021COSX3}

\section{Computational details}
To expedite subsequent discussions, ERIJ and ERIK are used to denote analytical evaluation of 
the full $\mathbf{J}[\mathbf{D}]$ and $\mathbf{K}[\mathbf{D}]$ matrices, respectively.
MECP refers to analytic evaluation of $\mathbf{J}[\mathbf{D}^A_{\oplus}]$ and 
numerical evaluation of $\mathbf{J}[\mathbf{D}-\mathbf{D}^A_{\oplus}]$, whereas COSx 
refers here to analytic evaluation of $\mathbf{K}[\mathbf{D}^A_{\oplus}]$ and
semi-numerical evaluation of $\mathbf{K}[\mathbf{D}-\mathbf{D}^A_{\oplus}]$ with the present construction of S-junctions.
aMECP amounts to updating the $\mathbf{J}[\mathbf{D}^{RC}]$ portion [see Eq. \eqref{J-AOpair}] of $\mathbf{J}[\mathbf{D}-\mathbf{D}^A_{\oplus}]$
with semi-numerical integration upon convergence of the SCF iterations. 
Similarly, aCOSx amounts to updating the $\mathbf{K}[\tilde{\textbf{D}}^{RC}]$ portion [see Eq. \eqref{eqn:DRC}] of 
$\mathbf{K}[\mathbf{D}-\mathbf{D}^A_{\oplus}]$ with analytic integration upon convergence of the SCF iterations. 
While MECP and aMECP can adopt the same medium, fine or ultra-fine grids\cite{nwchem2010} as used in density functional calculations,
COSx and aCOSx will adopt the $G_{002}$, $G_{003}$, and $G_{013}$ grids as described in Sec. \ref{aCOSXSec}. 
Additional computational details will be exposed along with the presentation of the results.
All calculations were performed with the BDF program package\cite{Liu1997-BDF,Liu2003-BDF,Liu2004-BDF,Liu2020-BDF}.

\section{Results and discussion}\label{Results}
\begin{figure}[htp]
\begin{tabular}{c}
\resizebox{0.45\textwidth}{!}{\includegraphics{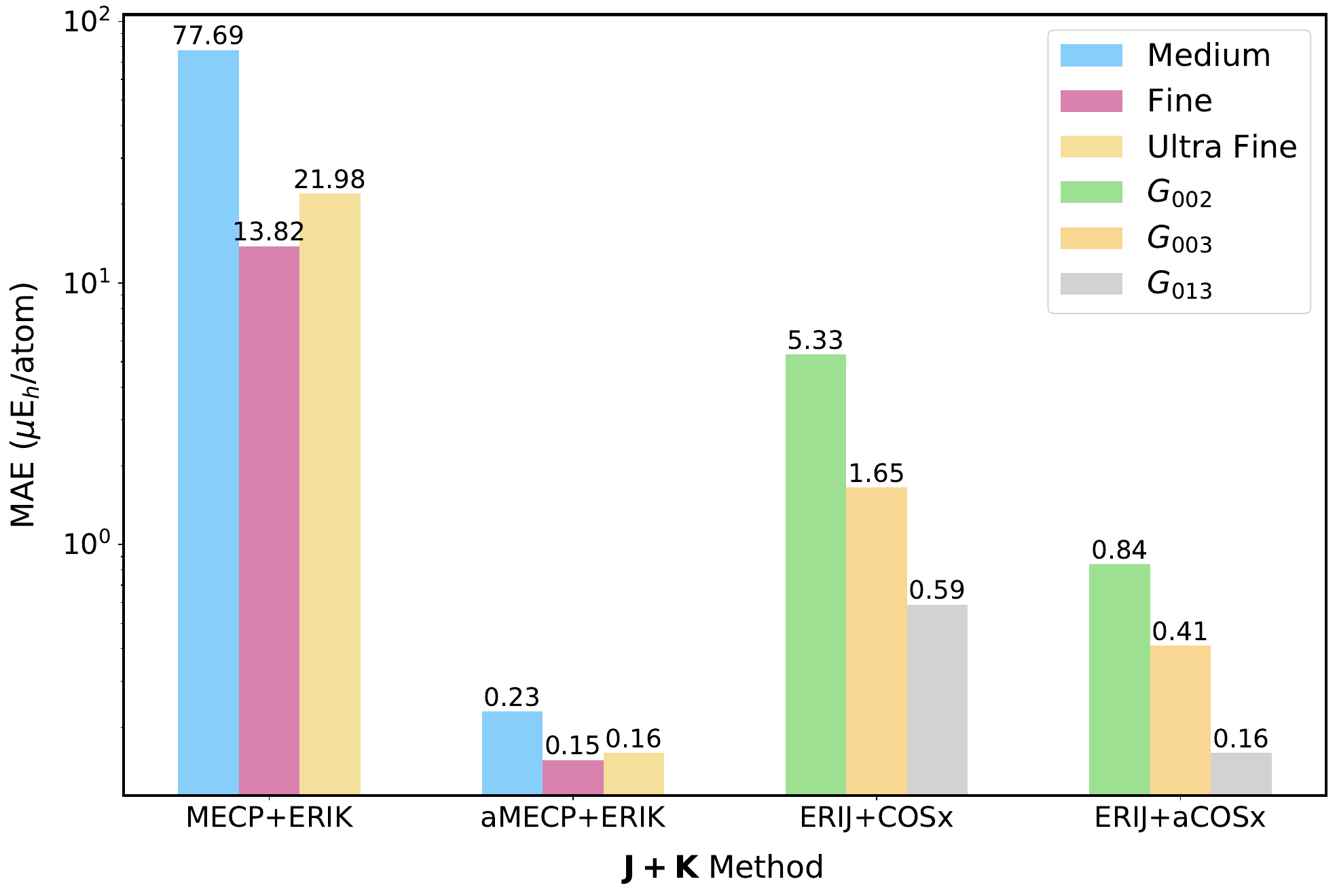}} \\
(a) \\
\resizebox{0.45\textwidth}{!}{\includegraphics{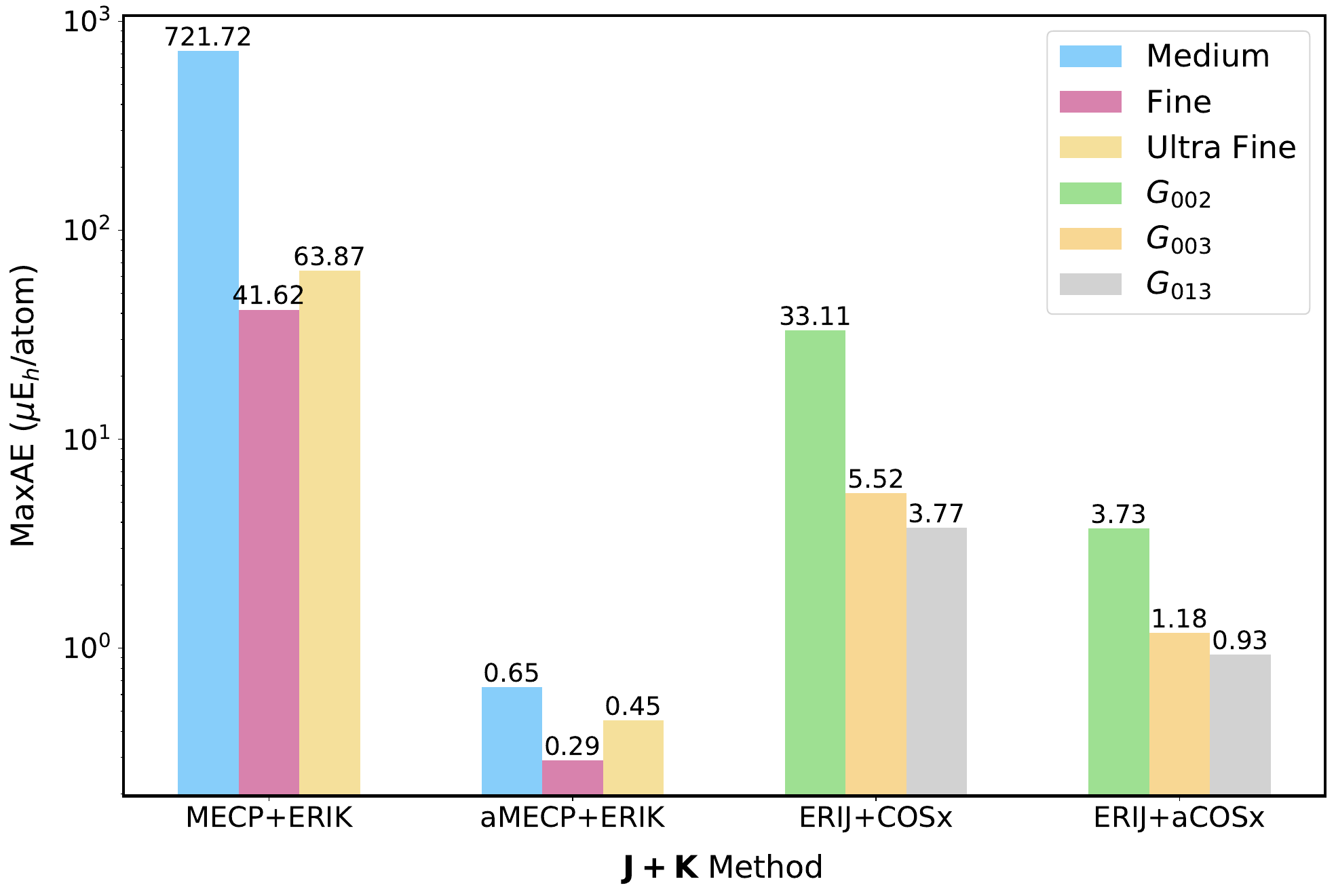}} \\
(b)
\end{tabular}
\caption{Errors of the MECP+ERIK, aMECP+ERIK, ERIJ+COSx, and ERIJ+aCOSx algorithms for the HF/cc-pVDZ total energies of Mole20. 
MAE: mean absolute error; MaxAE: maximum absolute error.}
	\label{fig:mole20HFvdz}
\end{figure}

To reveal the performances of the (a)MECP and (a)COSx algorithms, four algorithmic combinations, i.e., MECP+ERIK, aMECP+ERIK, ERIJ+COSx, and ERIJ+aCOSx, are first compared with ERIJK for the Hartree-Fock (HF) total energies of 20 molecules (denoted as Mole20) with the cc-pVDZ basis sets\cite{Dunning1989VDZ}.  
The Mole20 set includes 10 pharmaceutical molecules of the benchmark set used by Neese to access the COSX algorithm\cite{Neese2009COSX1}
and the following 10 molecules: two water clusters (\ce{(H2O)18} and Icecube27), a DNA molecule, \ce{C60}, a carbon nanotube (CNT40), beta-carotene, a bacteria-chromophore molecule (Bacl), a thermally activated delayed fluorescence molecule (TADFmol), the \ce{(amylose)4} chain, and olestra. Their 
Cartesian coordinates are provided in the Supporting Information. The mean absolute errors (MAE) and maximum absolute errors (MaxAE) of the methods are depicted in Fig. \ref{fig:mole20HFvdz}a and  \ref{fig:mole20HFvdz}b, respectively (see Tables S1 and S2 in the Supporting Information for more details).
It can be seen that, when the medium grid is employed in MECP, MECP+ERIK has a substantial MAE of 77.69 $\mu\mathrm{E}_h/\mathrm{atom}$,  which is accompanied by a MaxAE of 721.72 $\mu\mathrm{E}_h/\mathrm{atom}$.
When the fine grid is used, both the MAE and MaxAE are reduced but which cannot further be reduced with increased grid size. 
In contrast, the MAE of aMECP+ERIK falls below 1 $\mu\mathrm{E}_h/\mathrm{atom}$ and is not sensitive to grid sizes (e.g., 0.23, 0.15, and 0.16 $\mu\mathrm{E}_h/\mathrm{atom}$ for the medium, fine, and ultra-fine grids, respectively), thereby demonstrating the superiority of aMECP over MECP.  It is noteworthy that the MaxAEs of aMECP+ERIK utilizing the three different grids are also below 1  $\mu\mathrm{E}_h/\mathrm{atom}$, signifying the robustness of aMECP in  accuracy.
 
The errors of COSx and aCOSx can be revealed by comparing ERIJ+COSx and ERIJ+aCOSx with ERIJK. It can be seen from Fig. \ref{fig:mole20HFvdz}a and \ref{fig:mole20HFvdz}b
that the MAE of ERIJ+COSx with the small grid  $G_{002}$ is 5.33  $\mu\mathrm{E}_h/\mathrm{atom}$, while the MaxAE reaches 33.11 $\mu\mathrm{E}_h/\mathrm{atom}$. Increasing
the grid from $G_{002}$ to $G_{003}$ does reduce the error of COSx but at the expense of an increased computational cost. 
In contrast, even with the small grid $G_{002}$, 
the MAE of aCOSx is already below 1 $\mu\mathrm{E}_h/\mathrm{atom}$. In particular,
its MaxAE  is only 3.22  $\mu\mathrm{E}_h/\mathrm{atom}$, an order of magnitude lower than that of COSx.

\begin{figure}[htp]
\begin{tabular}{c}
\resizebox{0.45\textwidth}{!}{\includegraphics{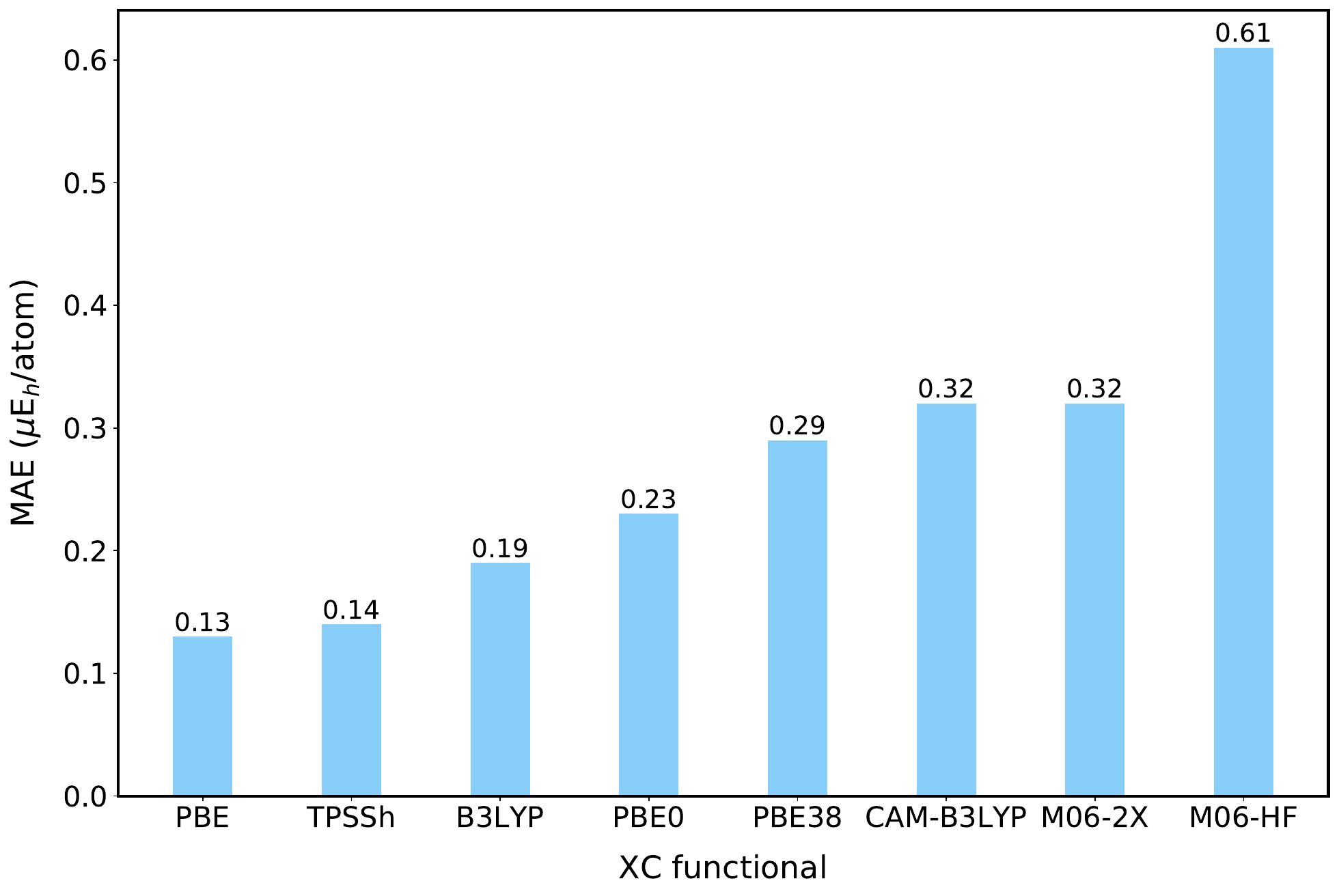}} \\
\end{tabular}
\caption{Mean absolute errors (MAE) of DFT calculations of Mole20 with aMECP+aCOSx and different functionals.}
	\label{fig:methodcompare}
\end{figure}

Now the question is how the combination of aMECP and aCOSx performs. It is known from 
Fig. \ref{fig:mole20HFvdz} that the error of aCOSx tends to be somewhat larger than that of aMECP for the Mole20 benchmark set. 
It can hence be anticipated that the MAE of aMECP+aCOSx is dominated by that of aCOSx. The latter should decrease
for hybrid functionals with less exact exchange. 
To confirm this, we took  M06-HF\cite{M06HF2006}, seven hybrid functionals (M06-2X\cite{M062X}. Cam-B3LYP\cite{CAMB3LYP}, PBE38\cite{PBE38},
PBE0\cite{PBE0}, B3LYP\cite{Becke93,B3LYP}, and TPSSh\cite{TPSSh}), and 
one pure density functional (PBE\cite{PBE}), which have decreasing percentage of exact exchange. 
All these functionals were implemented by using libxc.\cite{libxc2018}. The basis sets were 
def2-TZVP\cite{DEF2TVZP}. As can be seen from Fig. \ref{fig:methodcompare}, the MAEs of aMECP+aCOSx do decrease as decrease of the amount of 
exact exchange. 

\begin{figure}[htpb]
\begin{tabular}{c}
\resizebox{0.45\textwidth}{!}{\includegraphics{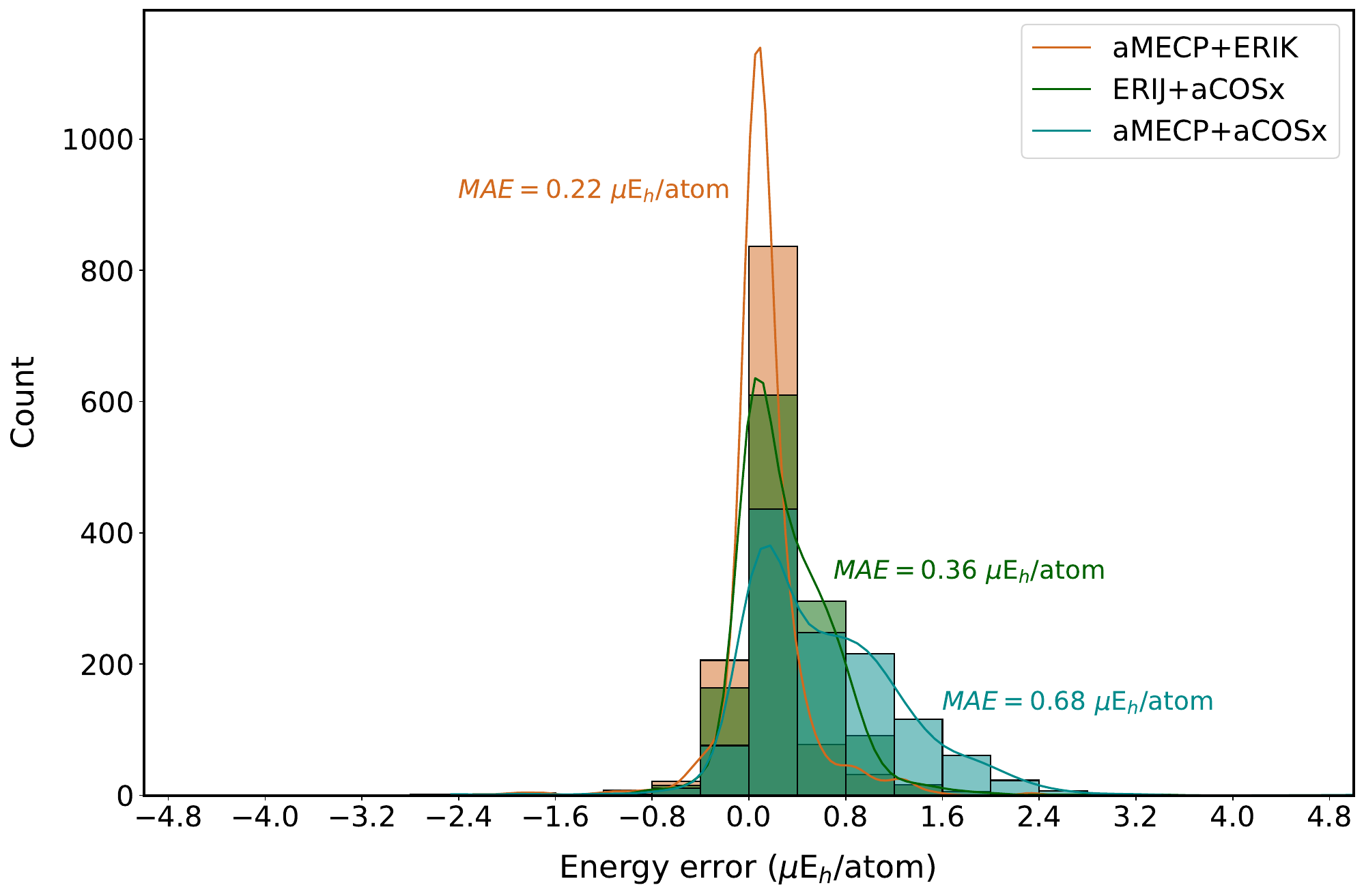}} \\  
(a) \\
\resizebox{0.45\textwidth}{!}{\includegraphics{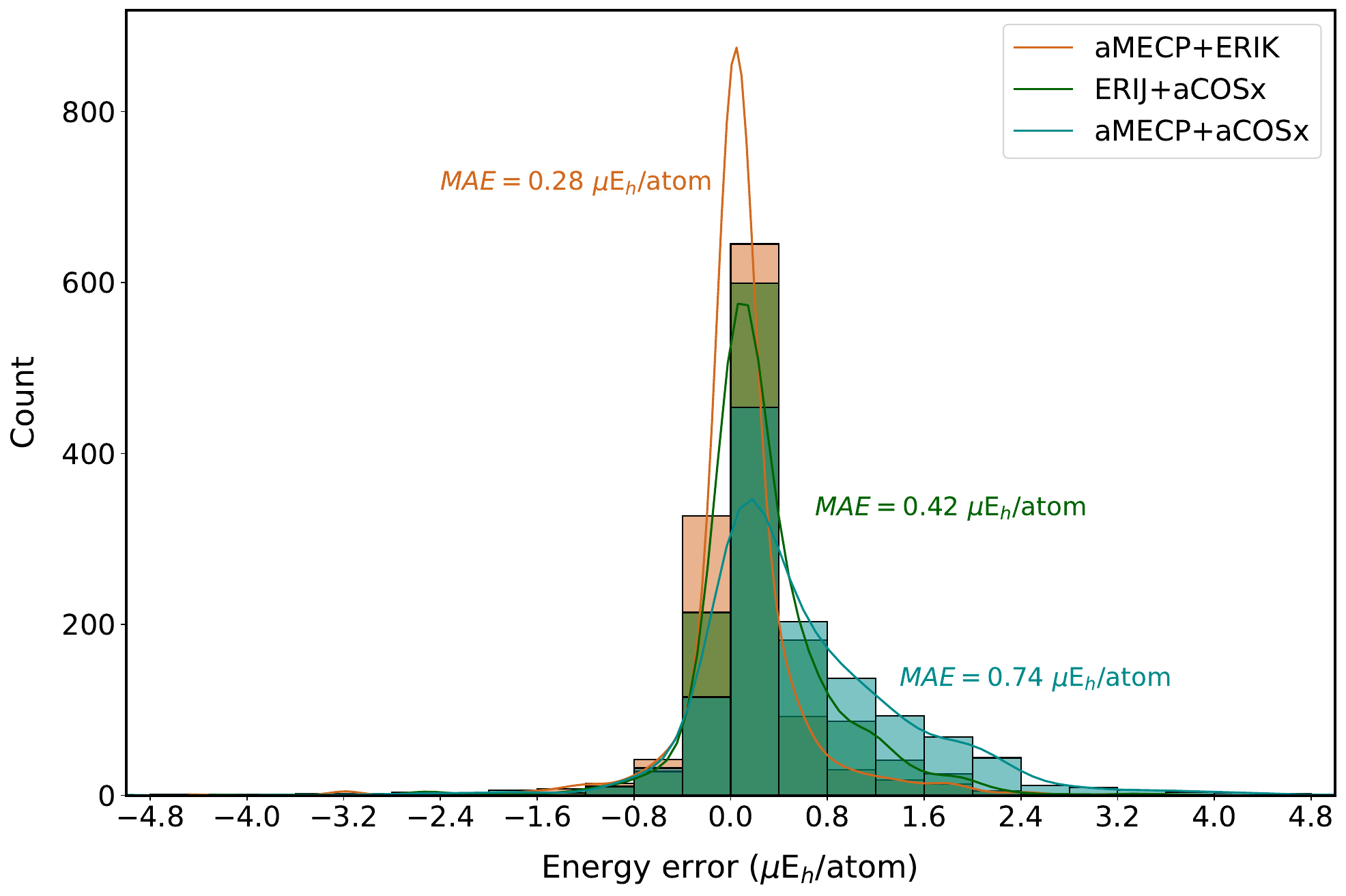}} \\
(b)\\
\resizebox{0.45\textwidth}{!}{\includegraphics{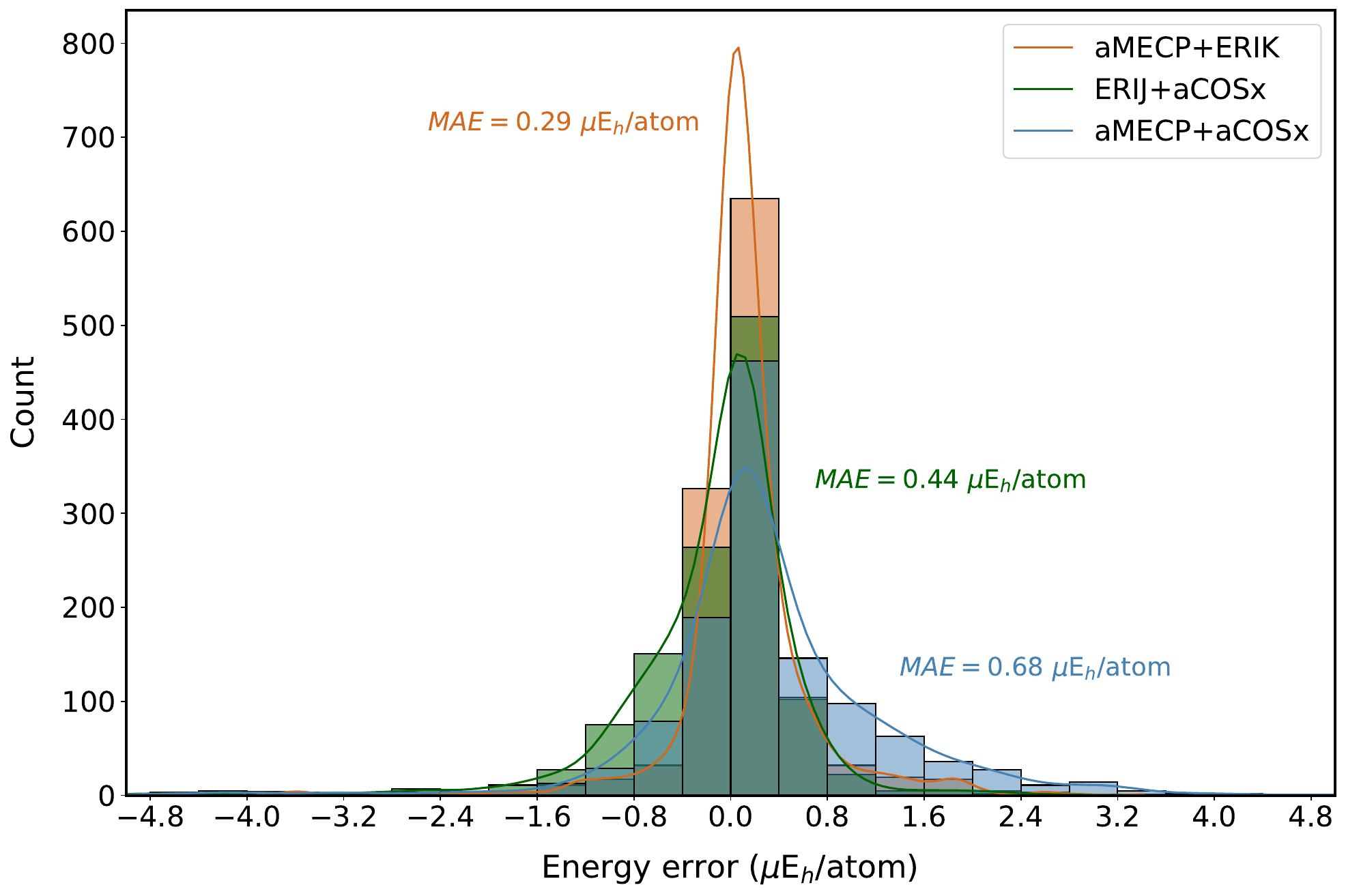}} \\
(c)\\
\end{tabular}
\caption{Distributions of energy errors as well as mean absolute errors (MAE) of aMECP+aCOSx, aMECP+ERIK, and ERIJ+aCOSx in HF 
calculations of 1210 molecules. (a) def2-SVP; (b) def2-TZVP; (c) def2-QZVP.}
	\label{fig:aMECPaCOSXBench}
\end{figure}

To have a more extensive examination of the aMECP+aCOSx-HF algorithm, we took 1210 molecules from the ACCDB database\cite{ACCDB2019} 
(see the Supporting Information for more details). The medium grid and the $G_{002}$ grid 
were used for aMECP and aCOSx, respectively. To check possible basis set dependence, 
three basis sets (def2-SVP, def2-TVZP, and def2-QVZP \cite{DEF2TVZP}) were employed. 
The distributions of the energy errors as well as the MAEs for aMECP+ERIK, ERIJ+aCOSx, aMECP+aCOSx 
are illustrated in Fig. \ref{fig:aMECPaCOSXBench}. It can be seen that 
aMECP+aCOSx is insensitive to basis sets, with the MAEs all 
below 1 $\mu\mathrm{E}_h/\mathrm{atom}$. Therefore, aMECP+aCOSx 
can be recommended as an efficient Fock build in HF and DFT calculations.

The accuracy of aMECP+aCOSx should be even higher for relative energies. To see this, we calculated the 
isomerization energies ($E_{\mathrm{iso}}$) of ISOL24 \cite{ISOL24} and reaction barriers ($E_{\mathrm{bar}}$) of a subset of INV24 \cite{INV24}. 
Again, the medium grid and the $G_{002}$ grid were used for aMECP and aCOSx, respectively, in conjunction with B3LYP/def2-TZVP. The
results are documented in Tables \ref{tbl:EISOCompare} and \ref{tbl:EBARCompare}, respectively. It can be seen that the 
MaxAEs of aMECP+aCOSx for E$_{\mathrm{iso}}$ and $E_{\mathrm{bar}}$ are only 0.022 and  0.056 kcal/mol, respectively.

\begin{figure}[h]
\begin{tabular}{c}
\resizebox{0.45\textwidth}{!}{\includegraphics{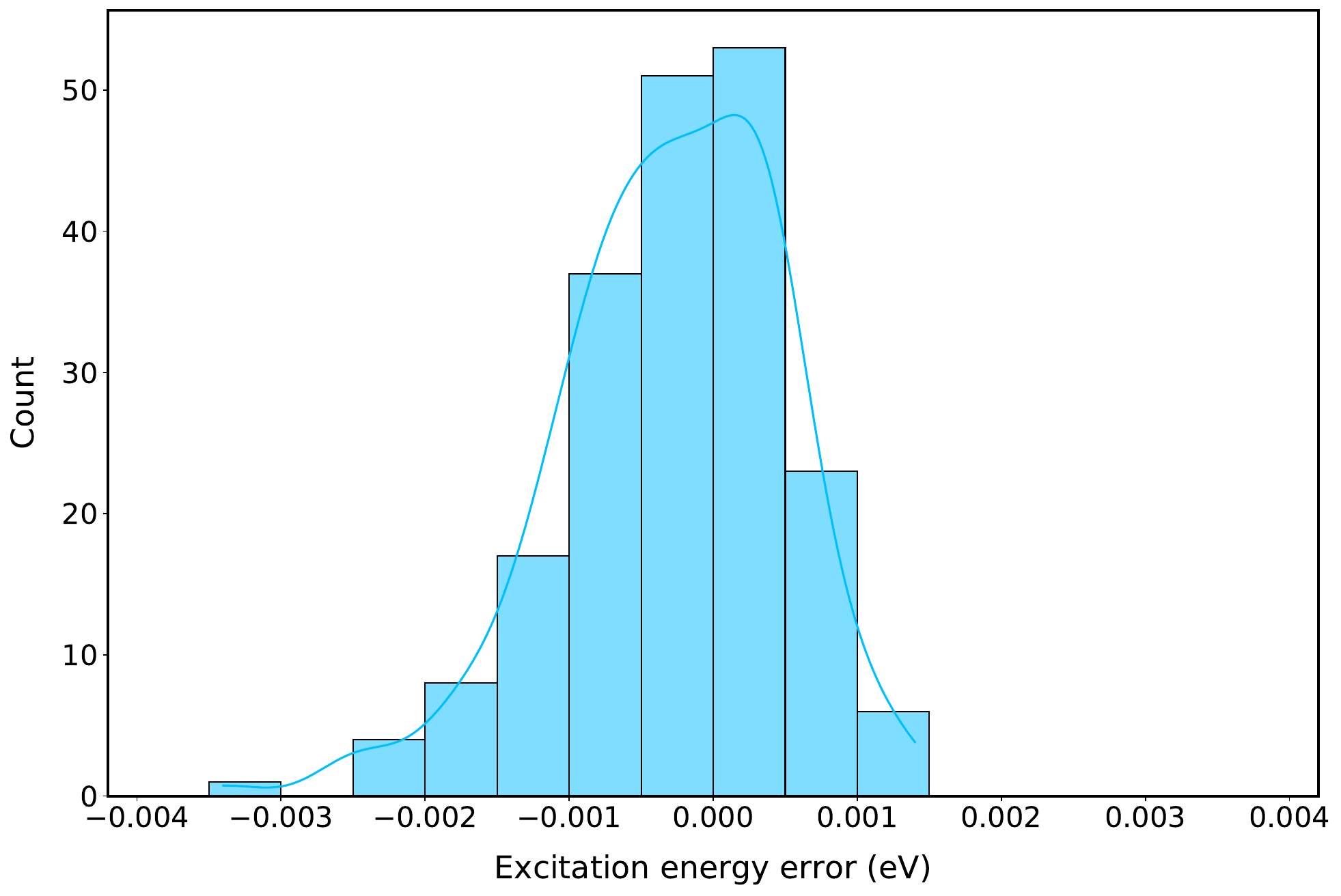}}
\end{tabular}
\caption{Histogram of errors for 200 TDDFT excitation energies with MECP+COSx.}
	\label{fig:tdenergy}
\end{figure}

MECP was known to be very accurate for TDDFT excitation energies\cite{BDFTDDFT2004,BDFTDDFT2005a,BDFTDDFT2005b}.
The question is how COSx performs in TDDFT calculations. To see this, we calculated 10 lowest singlet excited states 
for each molecule of the Mole20 set with MECP (medium grid) and COSx ($G_{002}$ grid), in conjunction with PBE0/cc-pVDZ.
The SCF calculations were performed by using aMECP+aCOSx.
 As can be seen from Fig. \ref{fig:tdenergy}, compared with the ERIJK-TDDFT calculations, the errors of the MECP+COSx-TDDFT excitation energies 
are typically within the range of -0.002 to 0.002 eV. Only four of the 200 excited states 
have errors somewhat larger than 0.002 eV, with the largest error being 0.004 eV. Moreover, the oscillator strengths
calculated by MECP+COSx are also very close to those by the ERIJK-TDDFT calculations (see Table S3 in the Supporting Information). 
Therefore, MECP+COSx can safely be used in TDDFT calculations after SCF calculations with aMECP+aCOSx. 

\begin{figure}[h]
\begin{tabular}{cc}
\resizebox{0.45\textwidth}{!}{\includegraphics{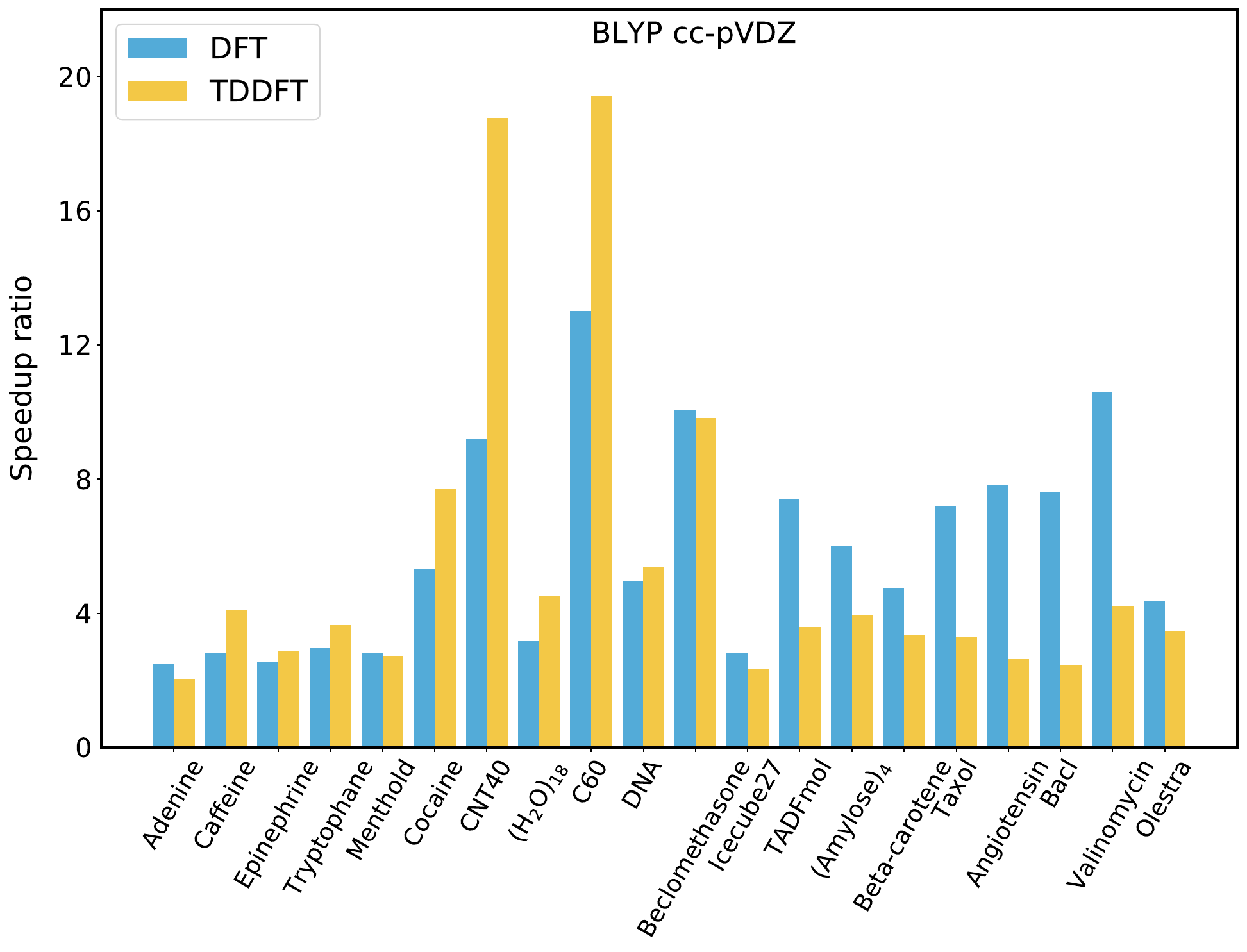}} &
\resizebox{0.45\textwidth}{!}{\includegraphics{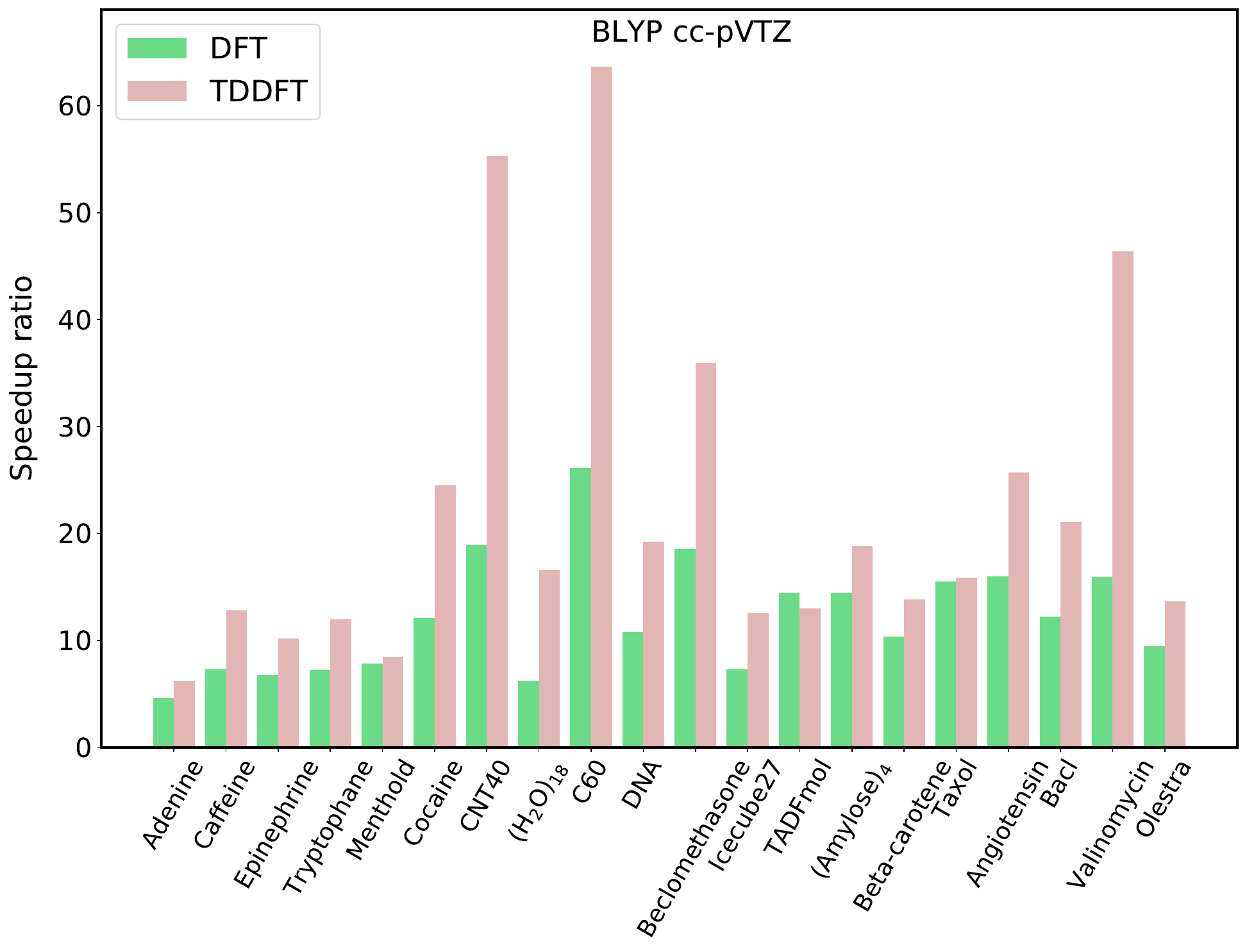}} \\
(a) & (b) \\
\resizebox{0.45\textwidth}{!}{\includegraphics{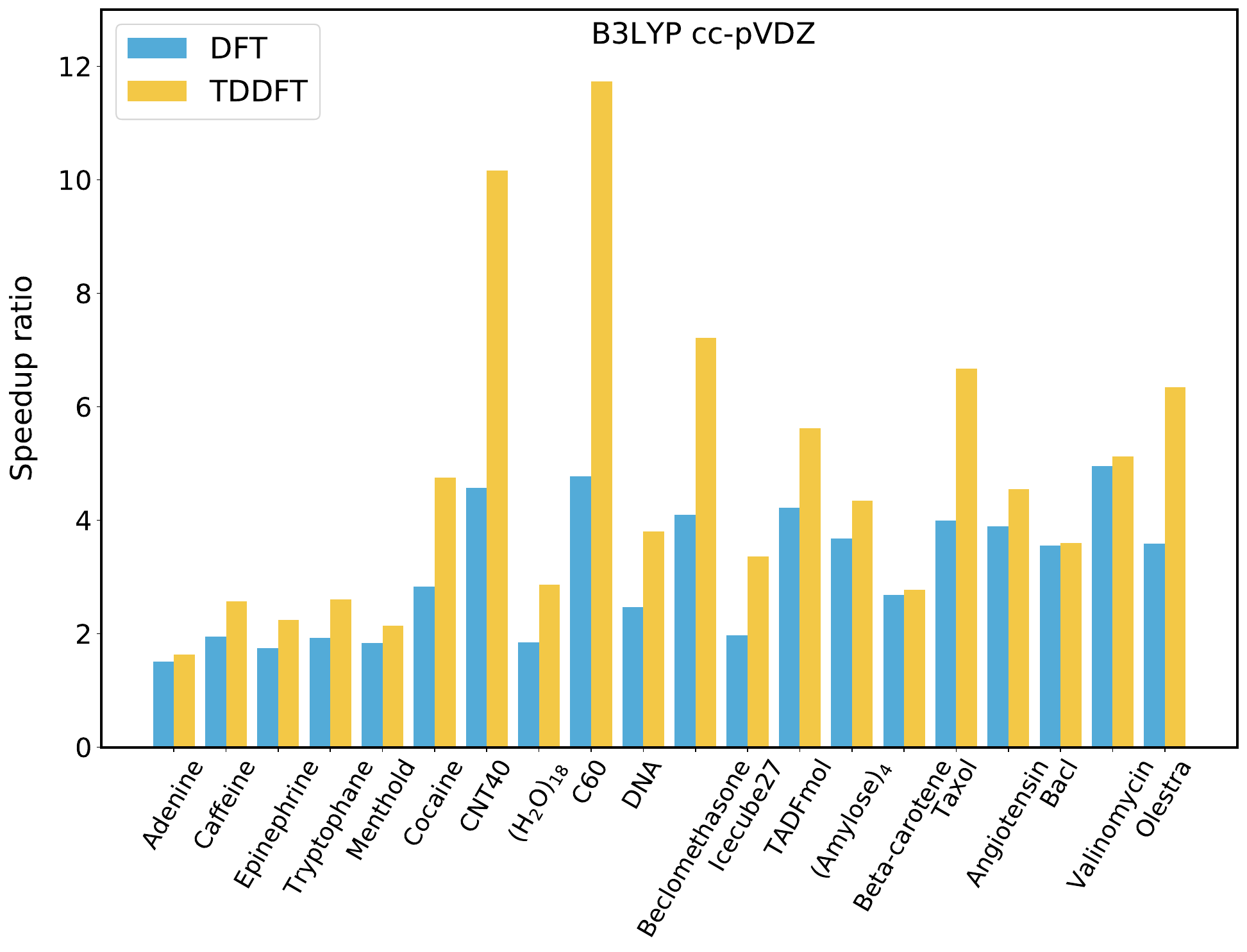}} &
\resizebox{0.45\textwidth}{!}{\includegraphics{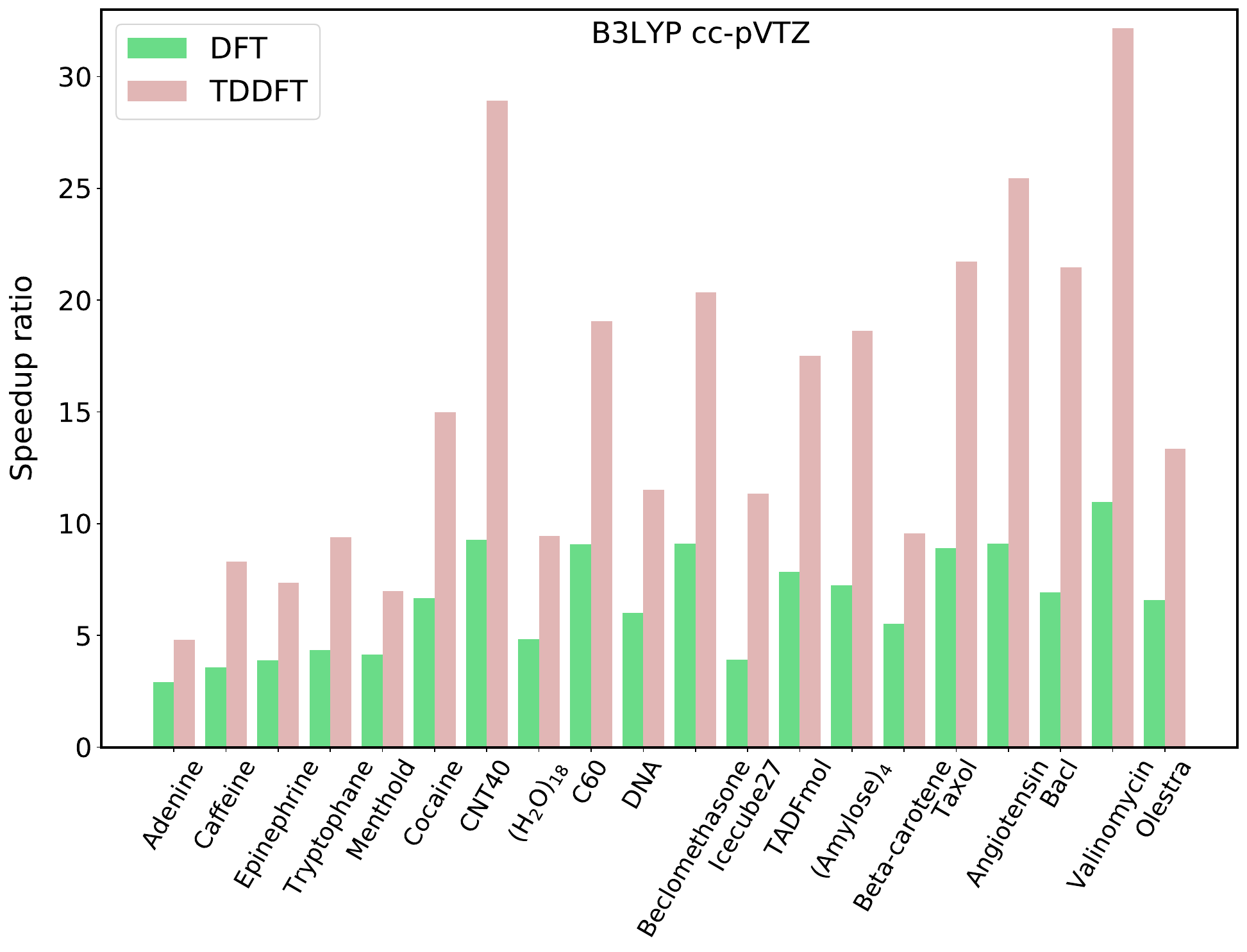}} \\
(c) & (d) \\
\end{tabular}
\caption{Speedups of (a)MECP+(a)COSx(e) over ERIJK in calculations of SCF energies and TDDFT 
excitation energies of Mole20. (a) BLYP/cc-pVDZ; (b) BLYP/cc-pVTZ;  (c) B3LYP/cc-pVDZ; (d) B3LYP/cc-pVTZ.}
	\label{fig:timespeedup}
\end{figure}

Having determined the accuracy of the aMECP+aCOSx and MECP+COSx algorithms for SCF and TDDFT calculations, respectively, 
it is time to check their speedups over ERIJK. It can be seen from Fig. \ref{fig:timespeedup} that 
the aMECP+aCOSx and MECP+COSx algorithms have achieved remarkable accelerations even for the smallest molecule, Adinane composed only of 15 atoms. 
The calculations with pure density functionals (e.g., BLYP) exhibit a more pronounced speedup over those with hybrid functionals (e.g., B3LYP).
This is not surprising, for the former involve only Coulomb-type ERIs. 
It is more gratifying to see that the fewer the hydrogen atoms, the larger the speedup (e.g, CNT40 and C60), and 
the more extended the basis set, the larger the speedup (e.g., 
the average speedups of 3.11 (SCF) and 4.70 (TDDFT) with cc-pVDZ vs. 6.54 (SCF) and 15.62 (TDDFT) with cc-pVTZ).   

\begin{figure}[h]
\begin{tabular}{c}
\resizebox{0.45\textwidth}{!}{\includegraphics{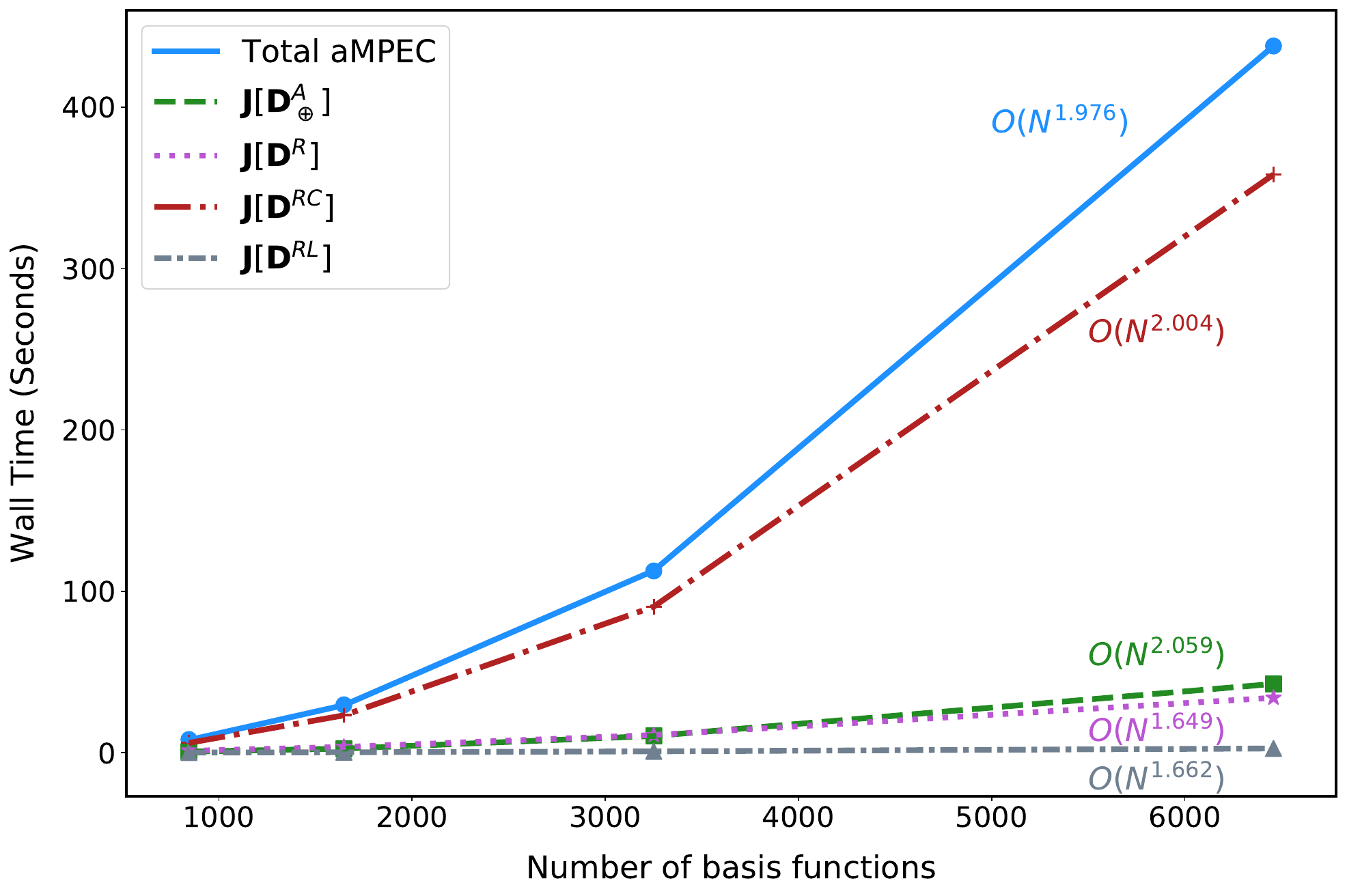}} \\
(a) \\
\resizebox{0.45\textwidth}{!}{\includegraphics{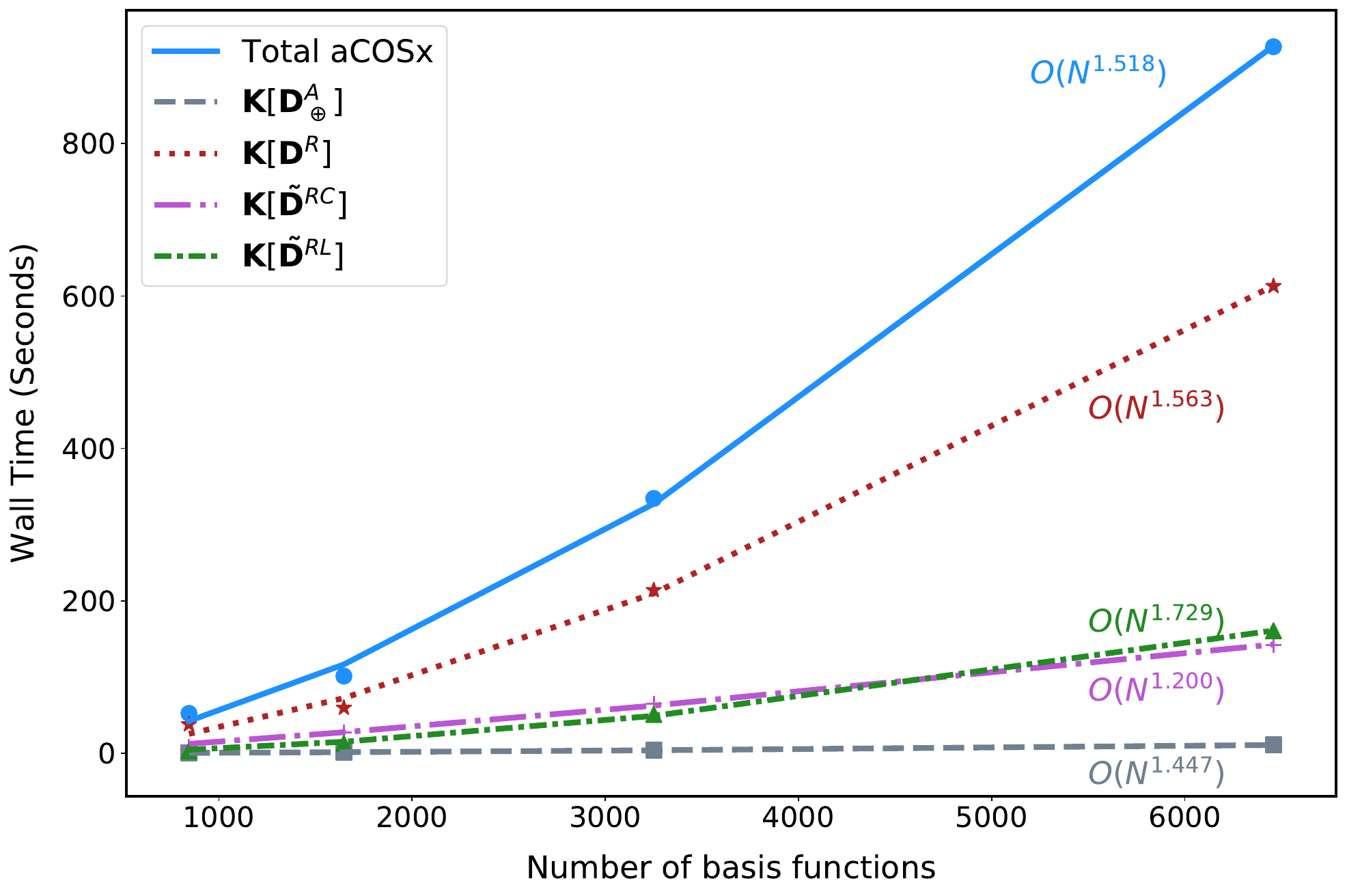}} \\
(b) \end{tabular}
\caption{Timings and scalings of aMECP+aCOSx-HF calculations of N-amylose chains, with 32 OpenMP threads
 on a computer equipped with an AMD 5975WX CPU and 64G memory.}
	\label{fig:methodscal}
\end{figure}

Finally, the timings and scalings of the $\mathbf{J}[\mathbf{D}^A_{\oplus}]$, $\mathbf{J}[\mathbf{D}^R]$, and $\mathbf{J}[\mathbf{D}^{RC}]$ and $\mathbf{J}[\mathbf{D}^{RL}]$  terms in aMECP as well as 
the $\mathbf{K}[\mathbf{D}^A_{\oplus}]$, $\mathbf{K}[\mathbf{D}^R]$, $\mathbf{K}[\tilde{\mathbf{D}}^{RL}]$, and $\mathbf{K}[\tilde{\mathbf{D}}^{RC}]$ terms in aCOSx are shown in Fig. \ref{fig:methodscal}.
The HF/def2-TVZP calculations were performed for the N-amylose chains (N = 2, 4, 8, 16). As can be seen from Fig. \ref{fig:methodscal}(a),
the $\mathbf{J}[\mathbf{D}^A_{\oplus}]$,  $\mathbf{J}[\mathbf{D}^R]$ and $\mathbf{J}[\mathbf{D}^{RL}]$  terms are very cheap, whereas the $\mathbf{J}[\mathbf{D}^{RC}]$ term consumes about 80\% of the aMECP runtime, although it is evaluated semi-numerically only once. 
The most expensive quantity, $A^g_{\lambda\mu}$ \eqref{eqn:amat}, involved in $\mathbf{J}[\mathbf{D}^{RC}]$ scales formally as $O(N_{B}^3)$
but actually as $O(N_B^{2.004})$, due to the fact that the number of AO pairs scales linearly with respect to
the molecular size (measured by the number ($N_B$) of basis functions). As a result, the overall scaling of aMECP amounts
to $O(N_{B}^{1.976})$. For comparison, the overall scaling of aCOSx, $O(N_B^{1.518})$, is somewhat lower.
Among the terms in aCOSx, $\mathbf{K}[\mathbf{D}^R]$ is most expensive ($O(N_B^{1.563})$) during the SCF iterations, which is followed by the one-step
$\mathbf{K}[\tilde{\mathbf{D}}^{RL}]$ ($O(N_B^{1.726})$) and $\mathbf{K}[\tilde{\mathbf{D}}^{RC}]$ ($O(N_B^{1.200})$) in the end of SCF iterations.
In particular, the near-linear scaling term $\mathbf{K}[\tilde{\mathbf{D}}^{RC}]$ is highly recommended, for it improves the accuracy of COSx significantly. 

In sum, the aMECP+aCOSx/MECP+COSx algorithm proposed here is highly robust and highly efficient in SCF/TDDFT calculations,
particularly when extended basis sets are to be used. 

\begin{table}[htp]
\caption{Isomerization energies ($E_{\mathrm{iso}}$ in kcal/mol) of ISOL24\cite{ISOL24}  calculated by B3LYP/def2-TVZP with ERIJK and aMECP+aCOSx.}
\begin{threeparttable}
\begin{tabular}{ccrc}
\hline
\hline
Molecule & Natom &   $E_{\mathrm{iso}}$ \tnote{a}   & $\Delta E_{\mathrm{iso}}$\tnote{b}  \\
\hline
 i1  &   72 &   -75.182     & 0.007  \\
 i2  &   41 &    -17.250     & 0.001   \\
 i3  &   24 &      -6.844   & -0.001  \\
 i4  &   81 &    -18.184     & 0.015   \\
 i5  &   32 &    -21.964    & -0.004 \\
 i6  &   48 &    -22.700   &  0.008  \\
 i7  &   51 &      -6.192     &  -0.003  \\
 i8  &   43 &    -51.278    &   0.016   \\
 i9  &   32 &    -17.832     & -0.003 \\
 i10 &   35 &      -2.027   &  0.003  \\
 i11 &   30 &    -36.955   & -0.004 \\
 i12 &   40 &      -0.570   &  -0.004 \\
 i13 &   26 &    -29.284   & -0.003 \\
 i14 &   26 &      -2.978   &  0.001  \\
 i15 &   42 &        1.785    & -0.004 \\
 i16 &   51 &    -23.579  & 0.013   \\
 i17 &   44 &      1.813     &  0.022  \\
 i18 &   39 &    -12.612  & 0.004  \\
 i19 &   36 &    -20.941 & -0.013  \\
 i20 &   28 &     -4.613  & 0.014    \\
 i21 &   36 &      4.243    &  -0.019 \\
 i22 &   44 &     16.745  & 0.013     \\
 i23 &   39 &     45.483  & 0.003    \\
 i24 &   52 &     -7.354  & -0.001  \\
 MAE      &      &     &  0.007  \\
MaxAE    &       &    & 0.022     \\
\hline
\end{tabular}
\begin{tablenotes}
  \item[a] Calculated by ERIJK.
  \item[b] Deviations of aMECP+aCOSx from ERIJK.
\end{tablenotes}
\end{threeparttable}
\label{tbl:EISOCompare}
\end{table}

\begin{table}[htp]
\caption{Reaction barriers ($E_{\mathrm{bar}}$ in kcal/mol) of the INV24 subset\cite{INV24} calculated by B3LYP/def2-TVZP with ERIJK and aMECP+aCOSx.}
\begin{threeparttable}
\begin{tabular}{ccrr}
\hline
\hline
Molecule              & Natom & $E_{\mathrm{bar}}$ \tnote{a}  & $\Delta E_{\mathrm{bar}}$ \tnote{b}  \\
\hline
BNCorannulene                &  30   &   5.073  &   0.000  \\
BNSumanene                    &  33   &  24.325 &   0.000  \\
Corannulene                      &  30   &   9.917  &    0.005  \\
Dibenzocarbazole            &  34   &   3.254  &  -0.001  \\
Dibenzocycloheptene   &  27    &   9.087  &  - 0.005   \\
Hexahelicene                   &  42    &  35.597 &   0.037  \\
Methinecyanine               &  40   &  12.042  &   0.007 \\
Pentahelicene                  &  36    &  23.297 &   -0.006  \\
PPh3                                    &  34   &  24.939 &  -0.024 \\
Sumanene                         &  33   &  18.342 &  -0.019  \\
Tetrabenzopyracylene  &  38   &   6.561  &   0.056  \\
Tetrahelicene                   &  30   &   3.829  &   0.006  \\
Triazasumanene             &  30   &  38.482 &   0.009 \\
Triindenotriphenylene  &  42    &  65.273 &   0.017  \\
MAE    &  &    &  0.014    \\
MaxAE    & & & 0.056 \\
\hline
\hline
\end{tabular}
\begin{tablenotes}
  \item[a] Calculated by ERIJK.
  \item[b] Deviations of aMECP+aCOSx from ERIJK.
\end{tablenotes}
\end{threeparttable}
\label{tbl:EBARCompare}
\end{table}

\section{Conclusion}\label{Conclusion}
A very robust and efficient hybrid analytic-numerical Fock build, aMECP+aCOSx, has been developed for accelerating HF/DFT calculations.
The essential idea is to extract those portions of the Fock matrix (i.e.,  those 
corresponding to the superposition of atomic density matrices and the density matrix spanned by the near AO pairs)
that can readily be evaluated analytically, so as to minimize
numerical noises arising from the semi-numerical and numerical integrations. 
As a result, the combination of aMECP with a medium grid as used in DFT calculations and aCOSx with a coarse grid ($G_{002}$)
is already sufficient to achieve an accuracy of less than 1 $\mu\mathrm{E}_h/\mathrm{atom}$ in total energies. 
The acceleration of aMECP+aCOSx over the analytic Fock build is already seen in calculations of small molecular systems and is more enhanced
in calculations of large molecules with extended basis sets. In contrast, the combination of MECP (with a medium grid) and COSx
(with a coarse grid), where only the part of the Fock matrix corresponding to the superposition of atomic density matrices
is treated analytically, is already accurate enough for TDDFT calculations of excitation energies and oscillator strengths,
for a good reason: it is the orbital products instead of individual orbitals that are the actual basis in TDDFT (and any post-SCF) calculations.
Both aMECP+aCOSx and MECP+COSx can readily be extended to the relativistic domain, where the acceleration is expected to be even higher. 
But before doing this, the GPU implementation should first be finished. Work along these directions is being carried out at our laboratory.

\begin{suppinfo}
Formulas for the three-center-one-electron Coulomb integrals are first presented in Sec. S1. 
The errors of the aMECP+aCOSx algorithm for the HF energies of the Mole20 set are documented in Tables S1 and S2,
whereas the corresponding errors of the MECP+COSx algorithm for the TDDFT excitation energies and oscillator strengths
are documented in Table S3. Benchmark sets used to assess aMPEC+aCOSx-HF with different basis sets are described in Sec. S4. 
The Cartesian coordinates of the Mole20 set are compiled in the file Mole20Coordinates.zip.
\end{suppinfo}

\begin{acknowledgement} 
The research of this work was supported by National Natural Science Foundation of China
(Project Nos. 22373057 and 21873077), Mount Tai Climbing Program of Shandong Province, and 
Double First-class University Construction Project of Northwest University.
\end{acknowledgement}

\clearpage
\newpage

\bibliography{fockmatrix.bib}

\end{document}